\documentclass[10pt,a4paper]{article}
\usepackage[utf8]{inputenc}
\usepackage{authblk}
\usepackage{acronym}
\usepackage{tabularx}
\usepackage{multirow}
\usepackage{multicol}
\usepackage{graphicx}
\usepackage{subfigure}
\usepackage{url}
\usepackage[colorlinks=true,
            linkcolor=black,
            anchorcolor=black,
            citecolor=black,
            filecolor=black,
            menucolor=black,
            runcolor=black,
            urlcolor=black]{hyperref}

\usepackage{amsmath}
\usepackage{amsfonts}
\usepackage{amssymb}
\usepackage{amsthm}

\newcommand{\colonequals}{\mathrel{\mathop:}=}

\usepackage{color}
\definecolor{darkblue}{rgb}{0,0,.6}
\definecolor{darkred}{rgb}{.6,0,0}
\definecolor{darkgreen}{rgb}{0,.6,0}
\definecolor{red}{rgb}{.98,0,0}

\usepackage{listings}

\title{TSSort -- Probabilistic Noise Resistant Sorting}
\date{(as of January, 2011)}

\author[1,2]{Jörn Hees\thanks{j\_hees at cs.uni-kl.de}}
\author[2]{Benjamin Adrian}
\author[2]{Ralf Biedert}
\author[2,3]{Thomas Roth-Berghofer}
\author[1,2]{Andreas Dengel}
\affil[1]{CS Department, University of Kaiserslautern, Germany}
\affil[2]{KM Group, DFKI GmbH, Kaiserslautern, Germany}
\affil[3]{Institute of CS, University of Hildesheim, Germany}

\begin{document}
\newcommand{\redline}[1]{\ignorespaces\begin{scriptsize}\begin{itemize}\setlength{\parskip}{0pt}%
	\setlength{\parsep}{0pt}\color{darkred}#1\end{itemize}\end{scriptsize}\ignorespaces}

\newcommand{\introduce}[1]{\emph{#1}\index{#1}}
\newcommand{\reuse}[1]{\emph{#1}\index{#1}}

\newcommand{\lastAccessed}{(last accessed: Jan.\ 23rd, 2011)}

\newcommand{\lO}{\mathcal{O}}

\maketitle

\begin{abstract}
In this paper we present TSSort, a probabilistic, noise resistant, quickly converging comparison sort algorithm based on Microsoft TrueSkill.
The algorithm combines TrueSkill's updating rules with a newly developed next item pair selection strategy, enabling it to beat standard sorting algorithms w.r.t. convergence speed and noise resistance, as shown in simulations.
TSSort is useful if comparisons of items are expensive or noisy, or if intermediate results shall be approximately ordered.
\end{abstract}

\section{Introduction}
Comparison sorts order lists of items based on an underlying binary operator (e.g., $\leq$).
Traditionally this comparison operator is assumed to be noiseless.
Nevertheless, there are many sorting scenarios in which this is not the case.

One such area is the ranking of players (or teams) of a game.
In this area players are the items to sort according to their skill.
As skill usually is not directly observable, it can only be deduced from individual performances of players in competition with other players.
A well known sub-area is the ranking of chess players.
In the case of pairwise competitions, the outcome of a single match between two players can be interpreted as the result of a noisy comparison between both.

Another area benefiting from noise resistant sorting algorithms is the field of Human Computation.
Human Computation (HC) is a design methodology which solves complex problems by division into (atomic) sub-problems which are easily solvable by humans and then computes a solution of the complex problem by merging the sub-results.
A common HC scenario is the collaborative sorting of a list of items according to a specified property by dividing the problem down to pairwise comparisons which are then outsourced to humans.
One subclass of HC are Games with a Purpose (GWAPs) \cite{Ahn2006}.
For example, Matchin \cite{HackerAhn2009Matchin} orders pictures according to niceness or BetterRelations \cite{Hees2010} ranks facts according to importance, both by eliciting pairwise user preferences.

In the mentioned areas, individual comparisons are \emph{noisy}.
This means, that for example an item $i$ can be ``smaller'' $\prec$ than another item  $j$ in 90 \%, $i \approx j$ in 1 \% and $i \succ j$ in 9 \% of the performed comparisons.
Such varying results of the same comparison operation impose a large problem for traditional comparison sorts, as they would in most cases lead to poor sorting results.

After presenting foundations from the related field of rating algorithms in the next section, we will discuss desired properties of our sorting algorithm in Section~\ref{sec:props}.
In the following sections we will describe TSSort and the newly developed next item pair selection strategy before presenting simulation results and a discussion thereof.

\section{Foundations}\label{sec:relatedWork}
As mentioned in the introduction, the sorting of items according to a noisy comparison operator is related to (chess) rating algorithms.

		One of the most famous chess rating systems is the so called Elo rating \cite{Elo1978ChessRanking}.
		In its original version\footnote{Nowadays the United States Chess Federation (USCF) and Fédération Internationale des Échecs (FIDE) both use variants of the original version using a logistic distribution which was found to better fit the winning chances of weaker players. Nevertheless, the underlying ideas and problems for this work remain the same.}, the idea behind Arpad Elo's rating system is to distinguish between the unobservable true skill $s_i$ of a player and the player's observable performances $p_i$.
		While the true skill is assumed to only change slowly over time, the observed performances can vary from one chess match to the next (e.g., a player might have a good or bad day).
		Elo assumed that the observable performances of a player $i$ are normally distributed around the players true skill: $p_i \sim \mathcal{N}(s_i,\beta^2)$.
		Furthermore, Elo simplified his model by assuming that each player had the same variance $\beta^2$ of performances.

		With these considerations we can calculate the probability that a player $i$ wins against an opponent $j$ if their true skills $s_i$ and $s_j$ are known \cite{Elo1978ChessRanking}.
%
		The Elo rating system specifies an updating rule (linearized) based on the comparison of the probability of a win of player $i$ over $j$ and the true outcome of the game:
		\begin{equation*}
			\Delta = \underbrace{\alpha \beta \sqrt{\pi}}_\text{$K$-factor} \left( \frac{y+1}{2} - P(p_i > p_j | s_i,s_j) \right) = \alpha \beta \sqrt{\pi} \left( \frac{y+1}{2} - \Phi\left(\frac{s_i - s_j}{\sqrt{2 \beta^2}}\right) \right)  ,
		\end{equation*}
		where $y = 1$ if player $i$ wins, $y = 0$ in case of a draw and $y = -1$ in case player $i$ looses against $j$.
		After the game, $s_i$ is updated by adding $\Delta$ and $s_j$ by subtracting $\Delta$.

		The reason behind this updating rule is to compare the expectation of the outcome to the actual result.
		If player $i$ outperforms the expectation, the bracket term is larger than 0, and $s_i$ will rise.
		If $i$ performs as good as expected, the bracket term is close to 0, hence $s_i$ remains nearly unchanged.
		If $i$ performs worse than expected, the bracket term is less than 0, and $s_i$ decreases.
%


		It is easy to apply the Elo rating system to arbitrary sorting problems.
		Each item in a given list is assigned a rating $s_i$, which is updated depending on the outcome of comparisons.
		Afterwards, the items are sorted according to their score.
		The underlying idea of modeling individual performances of one item as observances of a random variable which is normally distributed around the true rating $s_i$ of the item, accounts for noisy comparison operations.

		Nevertheless, the Elo rating system itself does not include a selection strategy for the next best item pair to compare, especially none, which fulfills the desired properties mentioned in Seciton~\ref{sec:props}.
		In the case of chess, the next match is usually determined by tournament rules or the players themselves.
		Furthermore, the Elo rating system uses some simplifications which have a negative impact when being used as a sorting algorithm, as we will see in the comparison in Section~\ref{sec:ranking:comparison}.


%
		In order to overcome the problems of the Elo rating system, a rating algorithm is needed which also models the remaining uncertainty of each item's true rating.
%
		One such universal rating algorithm is TrueSkill \cite{Herbrich2007}.
		Developed by Microsoft Research, it currently is the standard rating algorithm for ranking online players using Microsoft's Xbox Live console.
		TrueSkill includes many features which are out of the scope of this paper (such as calculating individual player ratings from multi-player games possibly including multiple teams).
		Hence, in this work we only use the innermost part of TrueSkill, namely the its updating rules without the factor graph.



%
		Similar to the original version of the Elo rating system, individual performances $p_i$ of a chess player $i$ (or in our case of a game item\footnote{In the following we will stick to the chess player example as it is more intuitive to understand. In the end we will switch back to the interpretation of items.}) are interpreted to be observations of a normally distributed random variable.
		In contrast to the Elo rating system, TrueSkill does not simplify by assuming that all players' performance variances are the same, but instead models a player's performance to be normally distributed around $\mu_i$ with standard deviation $\sigma_i$: $p_i \sim \mathcal{N}(\mu_i,\sigma_i^2)$.

		One main idea behind TrueSkill is that if two players $i,j$ with $(\mu_i, \sigma_i)$ and $(\mu_j, \sigma_j)$ play against each other, we can calculate the probability $P(p_i > p_j | (\mu_i, \sigma_i),(\mu_ji, \sigma_j))$ that $i$ wins against $j$.
		After the game we can compare the real result to our expectation.
%
		If our expectation is violated we should update our parameter estimates to better model the observation.
		In contrast to the Elo rating system, updating the estimates consists of updating 4 values: $(\mu_i,\sigma_i),(\mu_j,\sigma_j)$.
		Updating the $\mu_i$ estimate component takes the former uncertainty $\sigma_i$ into account (analog for $j$):
		If the system was already very certain about player~$i$'s rating (small $\sigma_i$), the value $\mu_i$ is only changed a bit.
		If the system was unsure about its former estimate of the player rating (large $\sigma_i$), the $\mu_i$ update can be much larger (for example for new players).
		As the system learns most from surprises, updates (decreases) of the $\sigma$ values in such a cases are stronger, but also depend on the uncertainty of the other player's estimate.

		In order to summarize both parameters of the player rating's estimate into one rating score value, TrueSkill chooses a pessimistic estimate of the true player's rating: $s_i = \mu_i - 3\sigma_i$.
		In terms of probabilities this means that with a 99 \% ($1-\Phi(-3)$) probability the true rating score of the player will be higher than $s_i$.
		For players with high uncertainty this means that $s_i$ is much lower than $\mu_i$, while for players with small uncertainty they can roughly be the same.

		Similar to the Elo rating system, TrueSkill can be used to order arbitrary lists of items, just that in this case each item is assigned two values.
		After each updating step the items are sorted according to their pessimistic rating estimate.

		Whenever the Elo rating system or TrueSkill are combined with a next best item pair selection strategy, we will call the resulting sorting algorithm \emph{EloSort} and \emph{TSSort} respectively.
		TrueSkill itself provides one such strategy, namely its matchmaking heuristic.
		Given the rating estimates of two players, we can calculate their draw probability \cite{Herbrich2007}:
		\begin{equation*}
			q_\text{draw}(\beta^2, \mu_i, \mu_j, \sigma_i, \sigma_j) = \sqrt{\frac{2\beta^2}{2\beta^2 + \sigma_i^2 + \sigma_j^2}}\cdot \exp\left(-\frac{(\mu_i-\mu_j)^2}{2(2\beta^2 + \sigma_i^2 + \sigma_j^2)}\right) \;
		\end{equation*}
		As we will see in Section~\ref{sec:nextItemPairSelStrat}, maximizing the draw probability is just one selection strategy, which especially focuses on balanced matches between players.
		It tries to maximize fun, but as we will see in Section~\ref{sec:ranking:comparison} does not completely account for the desired properties as listed in the next section.

\section{Desired Properties}\label{sec:props}
%
%
%
	As mentioned in the introduction, the proposed algorithm is useful in the field of Human Computation.
	The listed properties are motivated by an online collaborative sorting scenario, in which a server tries to order a list of items by outsourcing the atomic pairwise comparisons ``$i \prec j$'' to humans.

	\subsection{Minimized Waste of Comparisons}
		It is easy to imagine that a poor selection strategy of the next item pair to compare can lead to many unnecessary comparisons.
		This is similar to the different strategies in comparison sorts, ranging from simple ideas as in BubbleSort ($\lO(n^2)$) to more sophisticated algorithms such as QuickSort (avg.\ $\lO(n\,log(n))$) or MergeSort ($\lO(n\,log(n))$).
		As in Human Computation each wasted decision is a wasted moment of human time, the goal is to find a rating algorithm which, compared to a state of the art sorting algorithm, can rate facts without wasting a lot of comparisons. 

	\subsection{Noise Resistance}
		Obviously the first property is contradictory to our second goal of sorting despite noise.
		Many modern comparison sorts are parsimonious \cite[pp. 61ff]{Knuth1991AxiomsAndHulls}, which means they will not perform unnecessary comparisons.
		Put in another way, such algorithms will not even notice that something is wrong with the comparison operator.
		This is nice as it does not lead to contradictions within the algorithm, but is bad as the resulting order will most likely be incorrect.

		Hence we would like to have an algorithm which can cope with at least moderate noise levels.

%
%
%
	\subsection{Quick Convergence and Partial Sorting}
		Another desired property of a noise resistant sorting algorithm is quick convergence towards a well sorted list.

		This is a distinction from our first property and standard sorting algorithms as they optimize the final (overall) number of comparisons, so the number of comparisons until the list is perfectly sorted.

		In between, standard algorithms might have very different states such as MergeSort which merges a decreasing number of smaller sorted temporary lists.
		Would MergeSort be interrupted or stopped before it has finished its work, there would be a high chance that a top item is located somewhere on top of the second temporary list, which could be somewhere in ``the middle'' just before the last merge run.

		As humans suddenly might stop sorting a list or often want to use intermediate results as early as possible, it is desirable to have an intermediate list which is sorted as good as possible after each step.

\section{Probabilistic Sorting Algorithms}\label{sec:tssort}
	Especially the desired noise resistance listed in the previous section indicates that standard comparison sorts are insufficient for our kind of problem.
	As mentioned in Section~\ref{sec:relatedWork} sorting under noise is very related to (chess) rating systems.

	Nevertheless, not all of such rating systems are directly usable as what we call \emph{probabilistic sorting algorithms}, as some lack a strategy on how to select the next best item pair.
	Even if such a strategy is present, as is the case with TrueSkill, it might not lead to the best results w.r.t. the aforementioned desired properties.
%
%
%

	\subsection{Next Item Pair Selection Strategies}\label{sec:nextItemPairSelStrat}
		As TrueSkill allows to calculate the draw probability $q_\text{draw}$ of a match between two items $i, j$ (see Section~\ref{sec:relatedWork}), one idea is to select the item pair, which maximizes $q_\text{draw}$.
		This is TrueSkill's matchmaking heuristic, trying to achieve balanced games, matching players together who are likely to have a neck-and-neck race.
		Hence, one idea is to select the item pair with the \introduce{maximum draw probability}:
		\begin{equation*}
			(i,j) = \text{arg}\max_{(i,j)} \; q_\text{draw}(\beta^2, \mu_i, \mu_j, \sigma_i, \sigma_j)
		\end{equation*}

		Aside from the fact that calculating the maximum draw probability for all pairs is computationally costly, it is doubtful whether this minimizes the amount of matches required for sorting.
		For example two items with equal $\mu$ and very low $\sigma$ are likely to have a high draw probability, but preferring their selection over those of items with high $\sigma$ which still need to be ordered will most certainly not lead to a minimal amount of decisions.

		Following our own argumentation in this example, we propose another approach:
		Select those two items whose Gaussians $\mathcal{N}(\mu_i,\sigma_i), \mathcal{N}(\mu_j,\sigma_j)$ overlap most, preferring wide overlaps.

		Expressing this idea in terms of $2\sigma$-intervals (corresponding to the 95 \% confidence interval) greatly reduces its computational complexity.
		Let $a,b$ be the lower and upper boundary of $i$'s $2\sigma$ interval, and $c,d$ those of $j$:
		\begin{align*}
			a,b & \colonequals \mu_i \mp 2\sigma_i\\
			c,d & \colonequals \mu_j \mp 2\sigma_j\\
			\text{wOver}_{i,j} & \colonequals \frac{\min(b,d) - \max(a,c)}{\max(b,d) - \min(a,c)} \cdot \max(b-a, d-c)
		\end{align*}
		Intuitively speaking, we divide the length of the scoring interval overlapped by both $2\sigma$ intervals by the length of the scoring interval overlapped by at least one and then weight the result  with the length of the longer $2\sigma$ interval.
		Non overlapping $2\sigma$ intervals will lead to negative values, which is acceptable for our purpose.

		We can then calculate the next best item pair to be the one with the \introduce{maximum weighted overlap}\index{maximum overlap}:
		\begin{equation*}
			(i,j) = \text{arg}\max_{(i,j)} \;  \text{wOver}_{i,j}
		\end{equation*}

		Another simplification which greatly reduces the computation costs of this selection strategy consists of not considering all possible pairs but just such pairs of successive partners of items, when sorted by their rating scores $s_i$.
		We will call this selection strategy of a best next game item pair the \introduce{maximum partner weighted overlap}\index{maximum partner overlap}.

		All of the above can also be applied to the Elo rating system, but degenerate to the maximum partner weighted overlap selection strategy.
		Additionally as in this case the intervals all are of the same width, weighting is unnecessary, which is why we call it \introduce{maximum partner overlap} selection strategy in the following.

		A comparison of the proposed best next game item pair selection strategy is part of the next section.

\section{Comparison of Different Approaches}\label{sec:ranking:comparison}

\begin{figure}[bp]
		\centering
		\includegraphics[width=.47\textwidth]{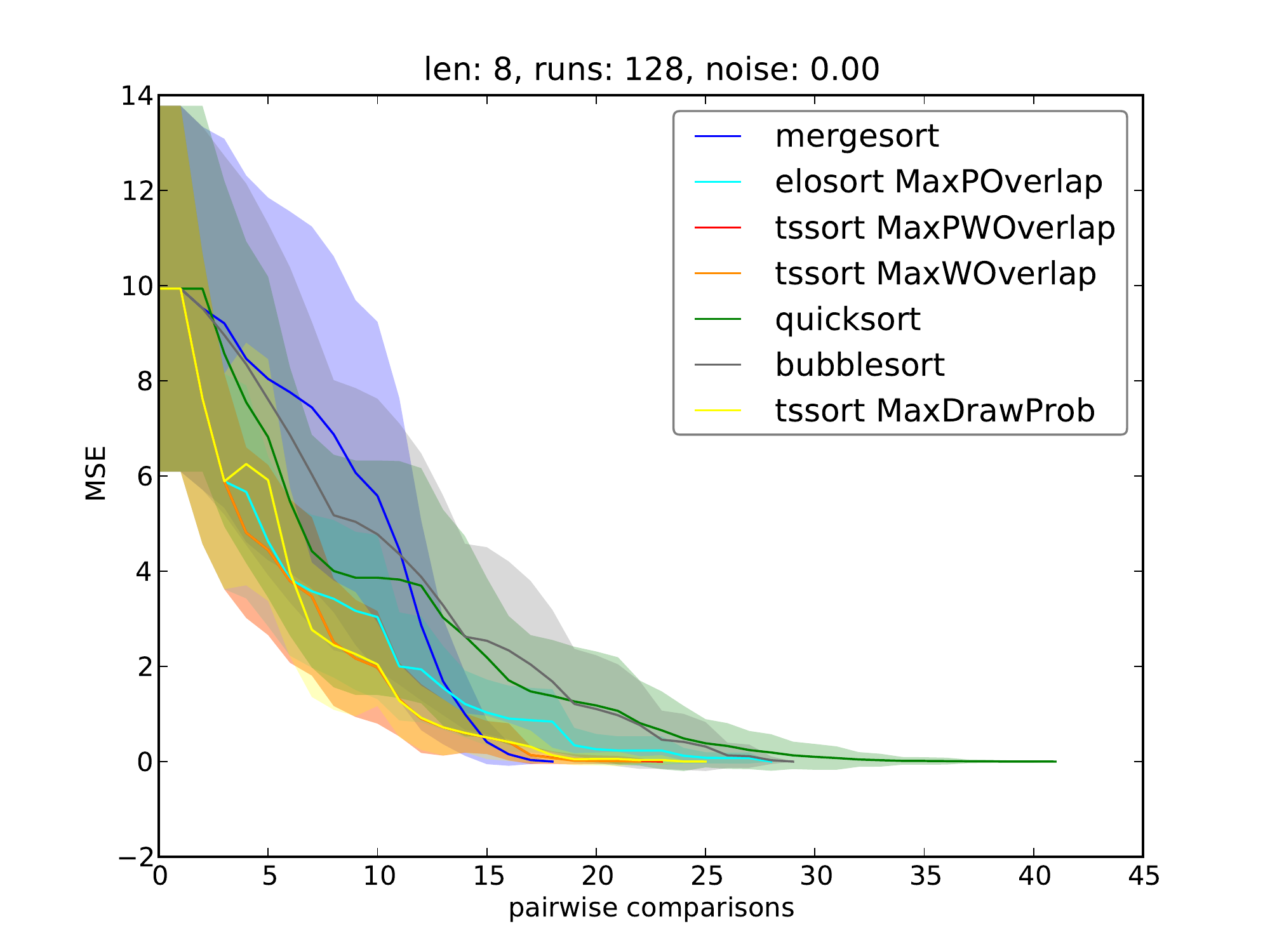} \hfill
		\includegraphics[width=.47\textwidth]{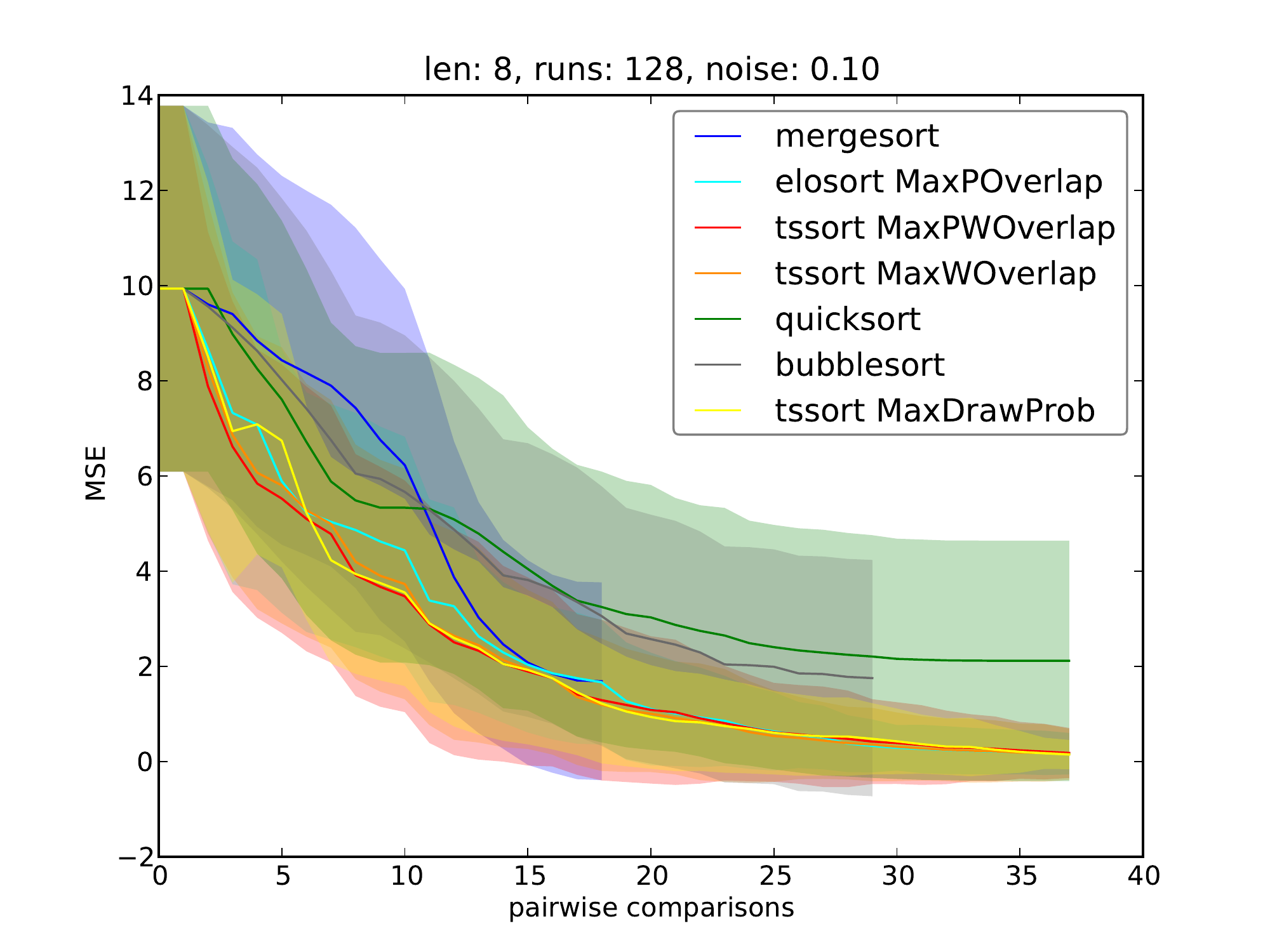} \hfill
		\includegraphics[width=.47\textwidth]{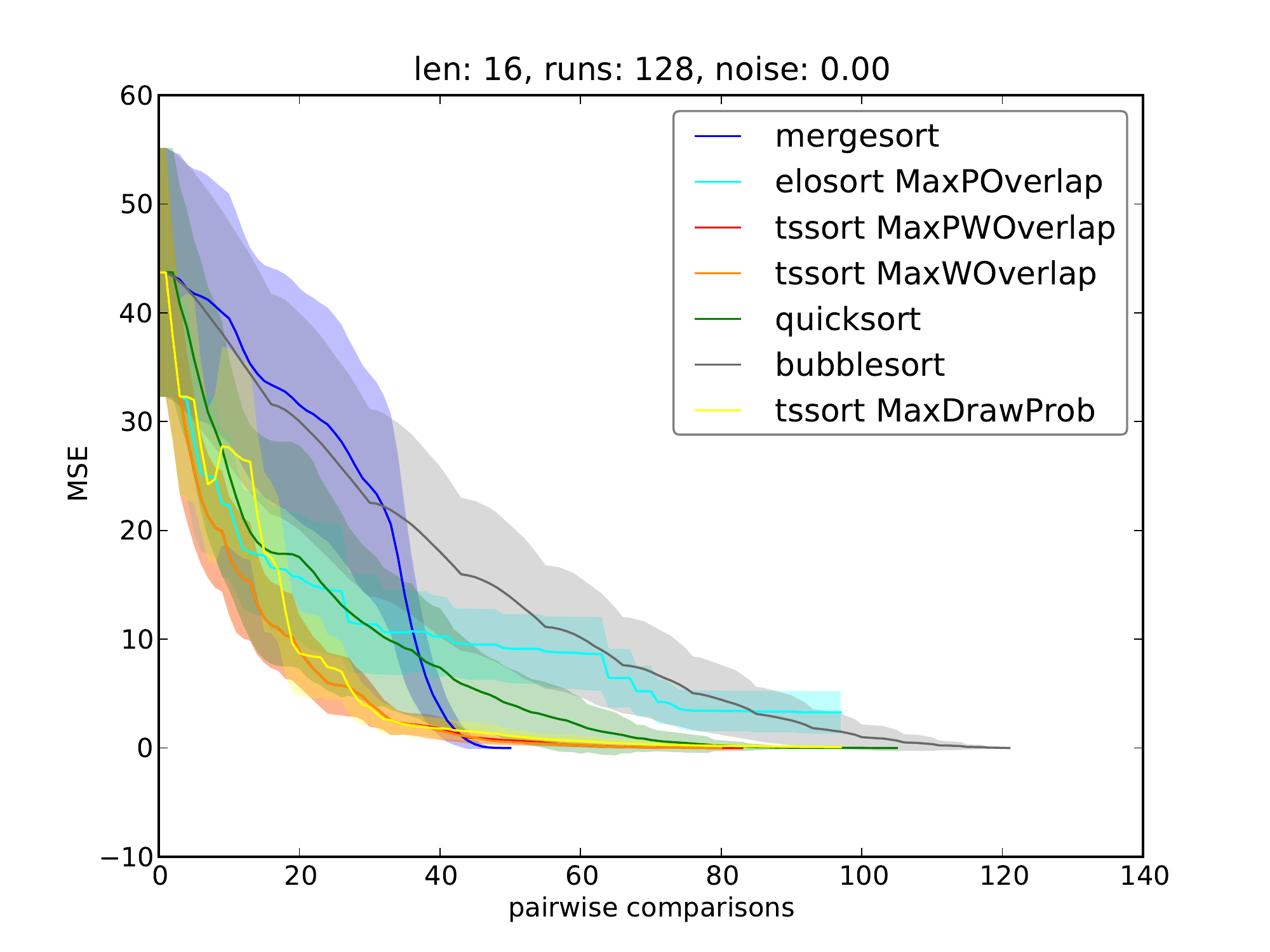} \hfill
		\includegraphics[width=.47\textwidth]{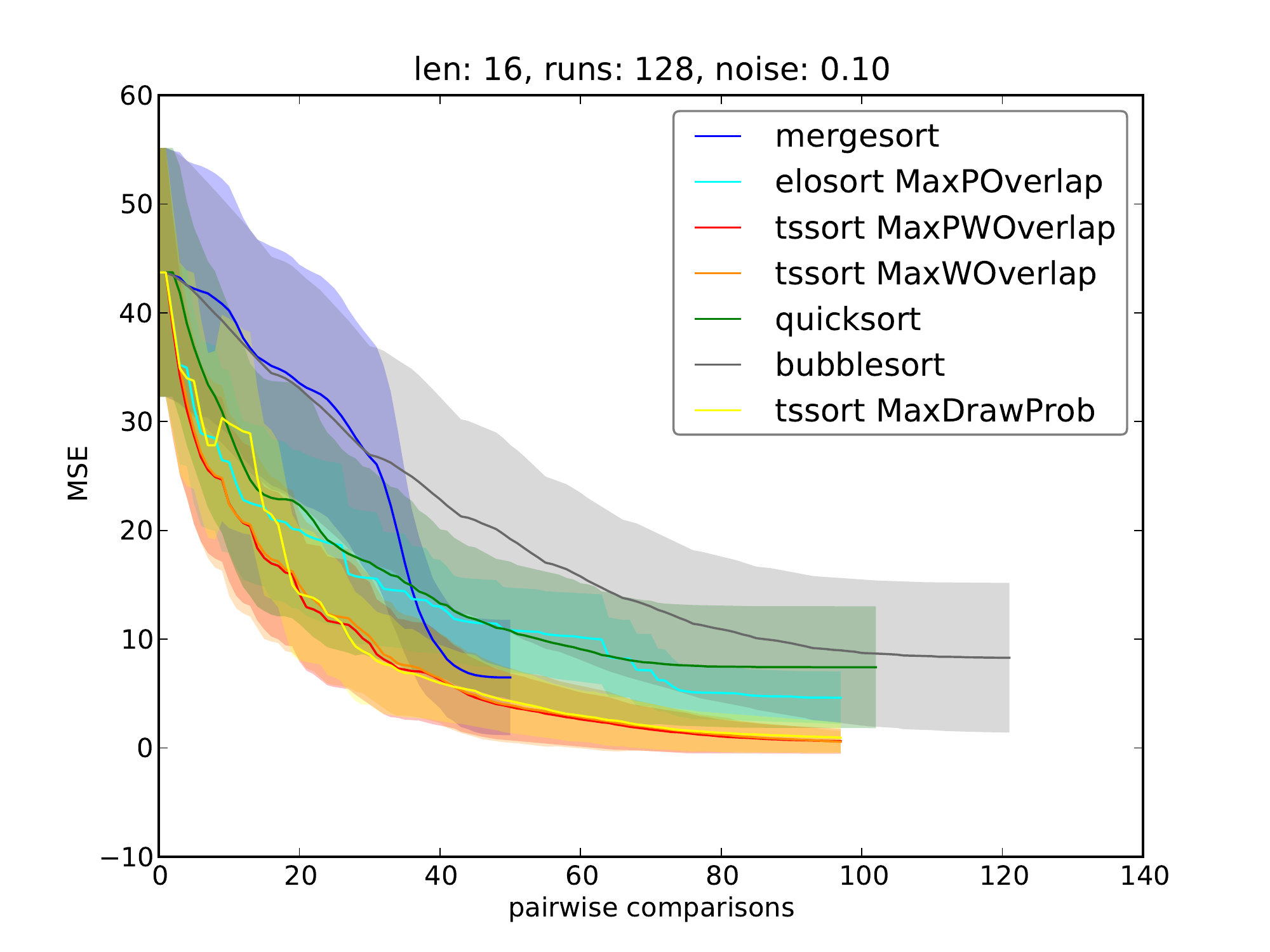} \hfill
		\includegraphics[width=.47\textwidth]{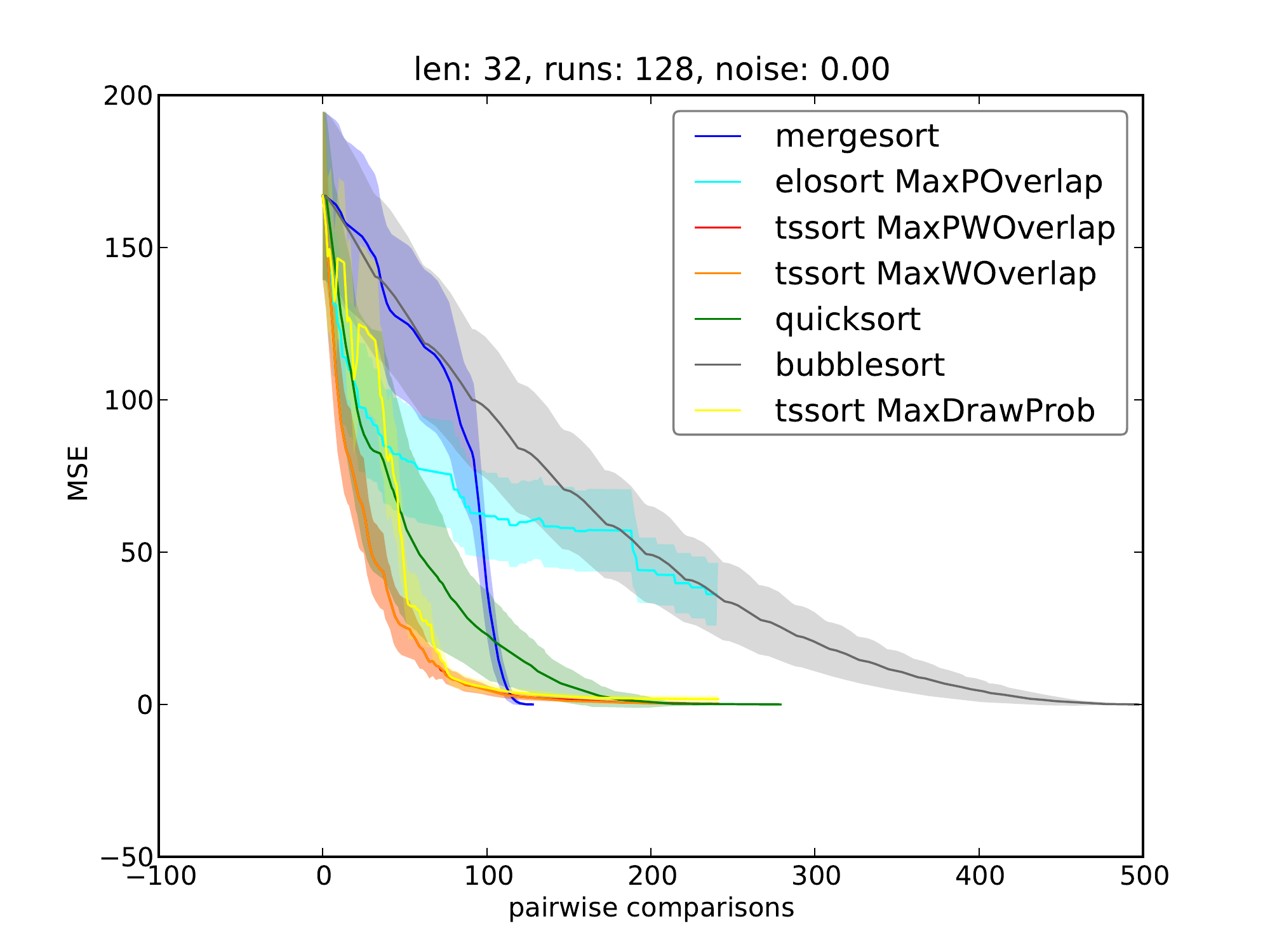} \hfill
		\includegraphics[width=.47\textwidth]{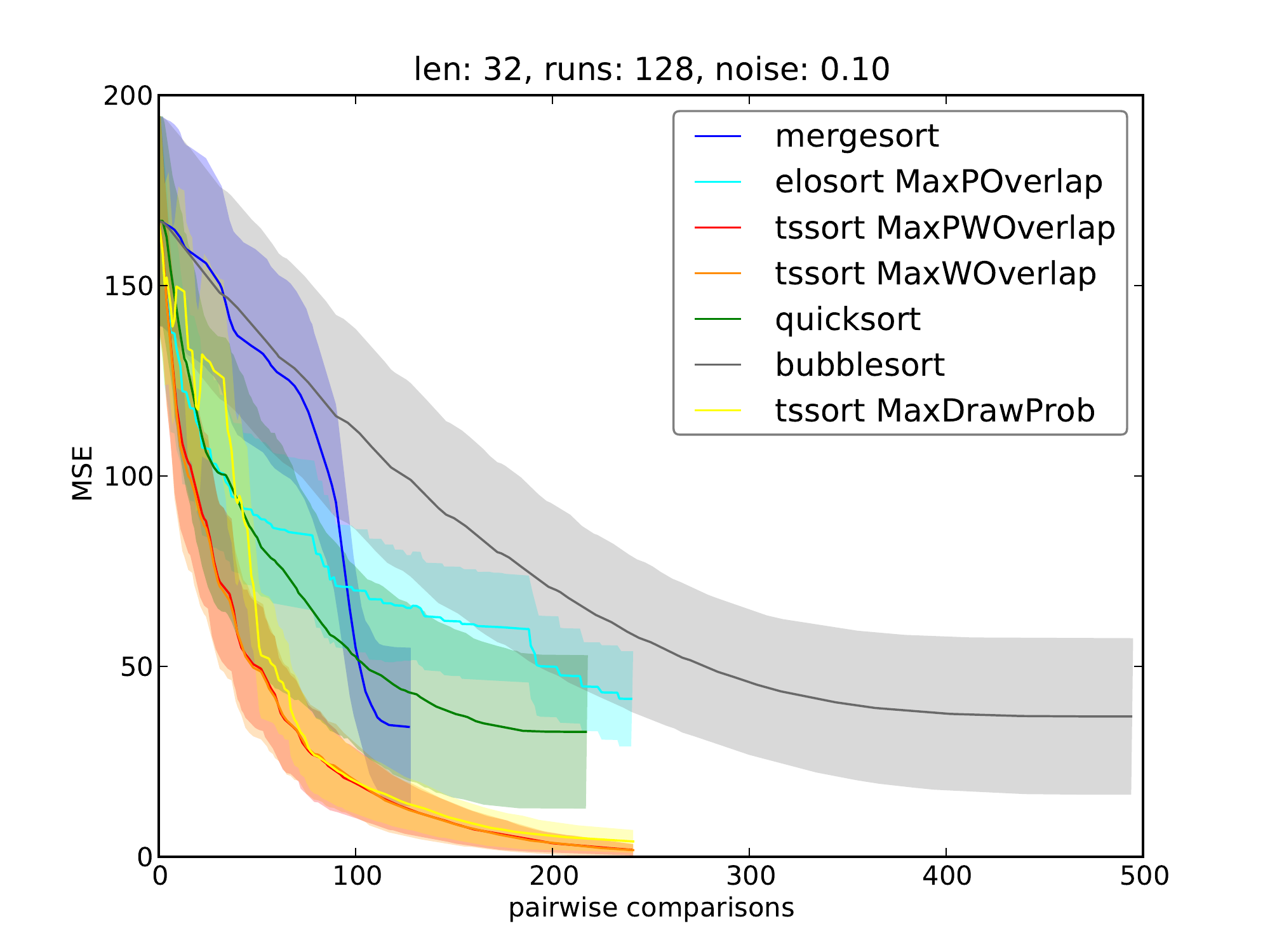} \hfill
		\includegraphics[width=.47\textwidth]{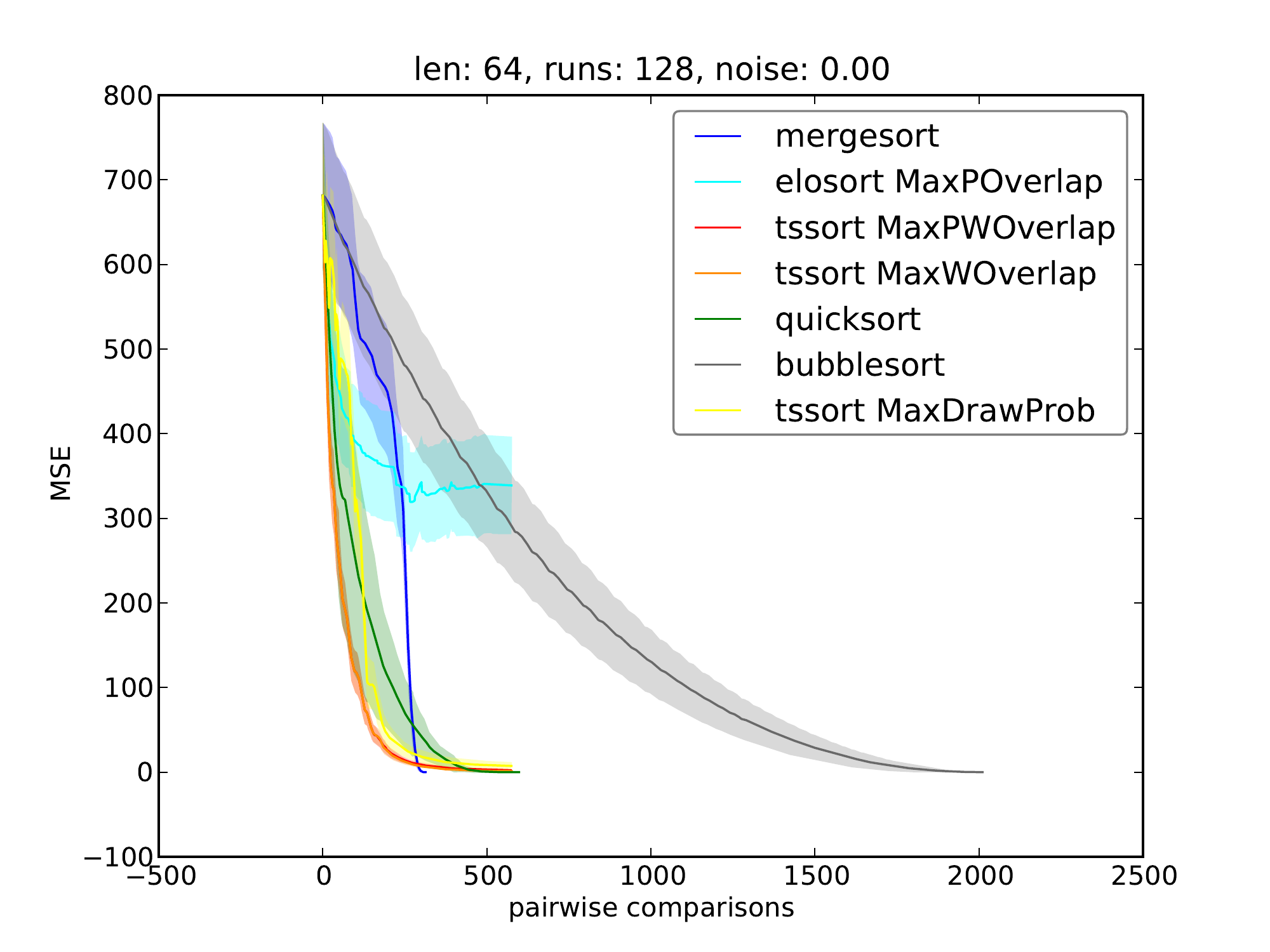} \hfill
		\includegraphics[width=.47\textwidth]{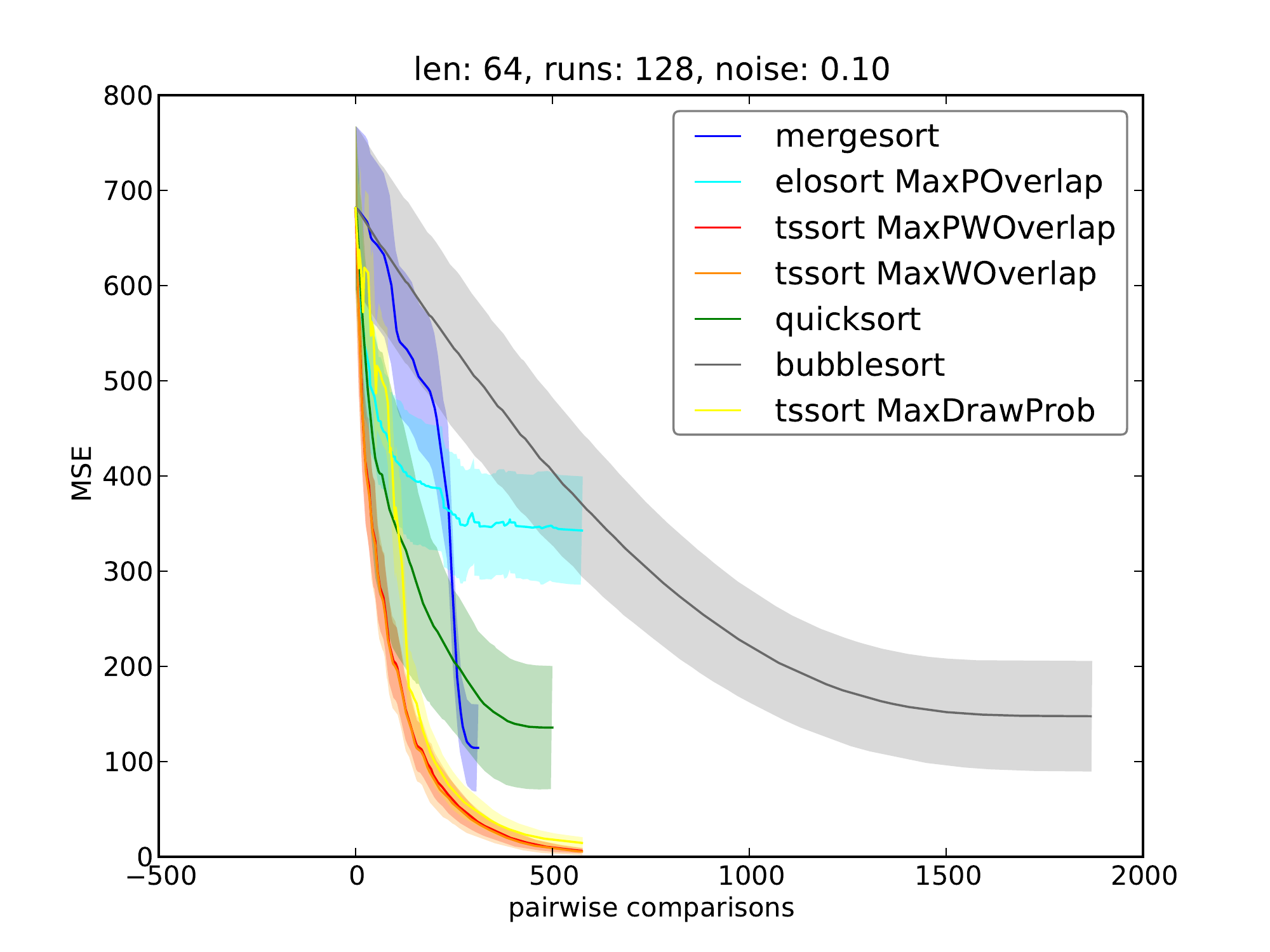} \hfill
		\caption{Comparison of sorting algorithms (Lists of length 8,16,32,64)}
		\label{fig:compSortAlgos1}
	\end{figure}

\begin{figure}[btp]
		\centering
		\includegraphics[width=.47\textwidth]{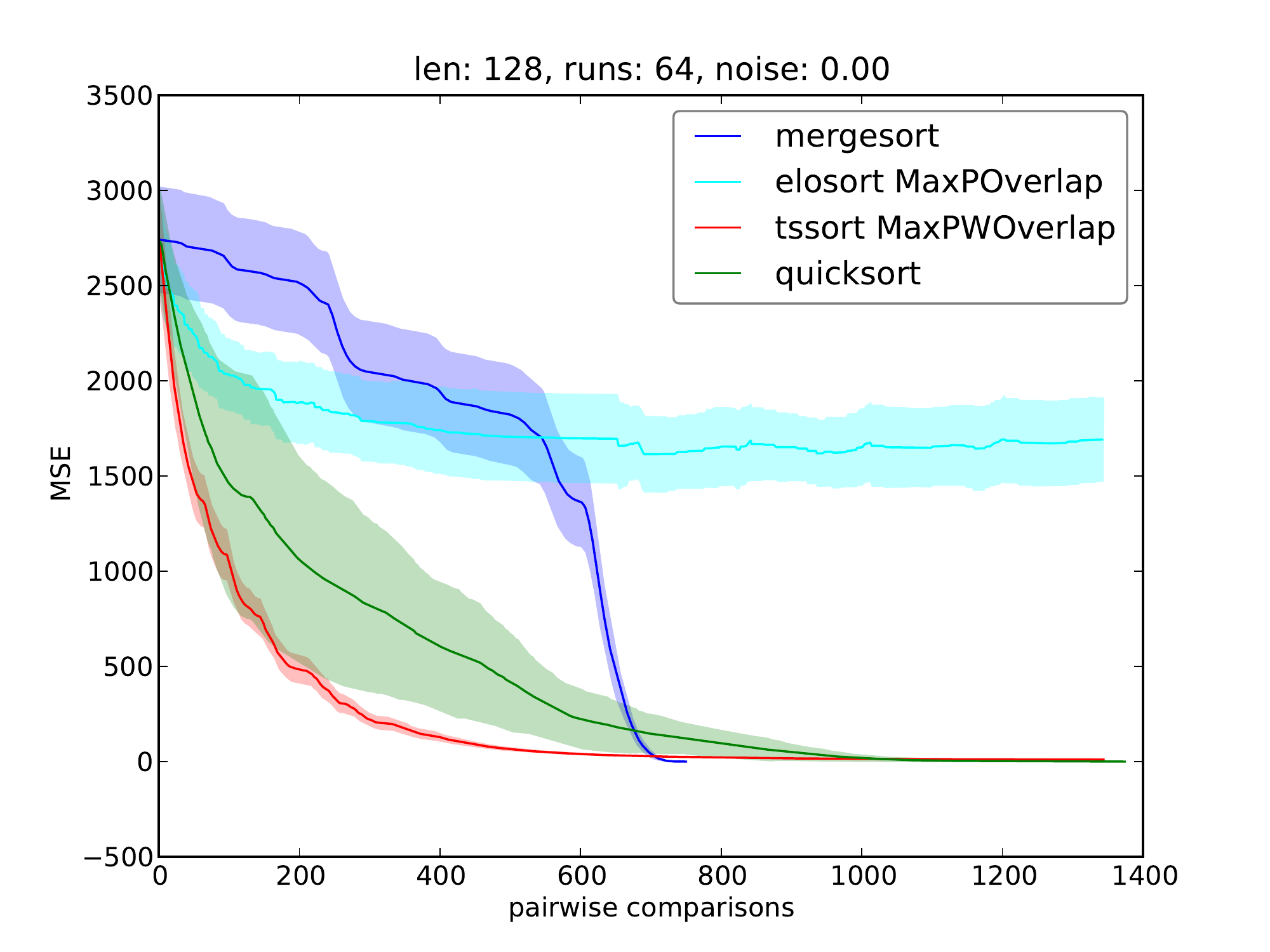} \hfill
		\includegraphics[width=.47\textwidth]{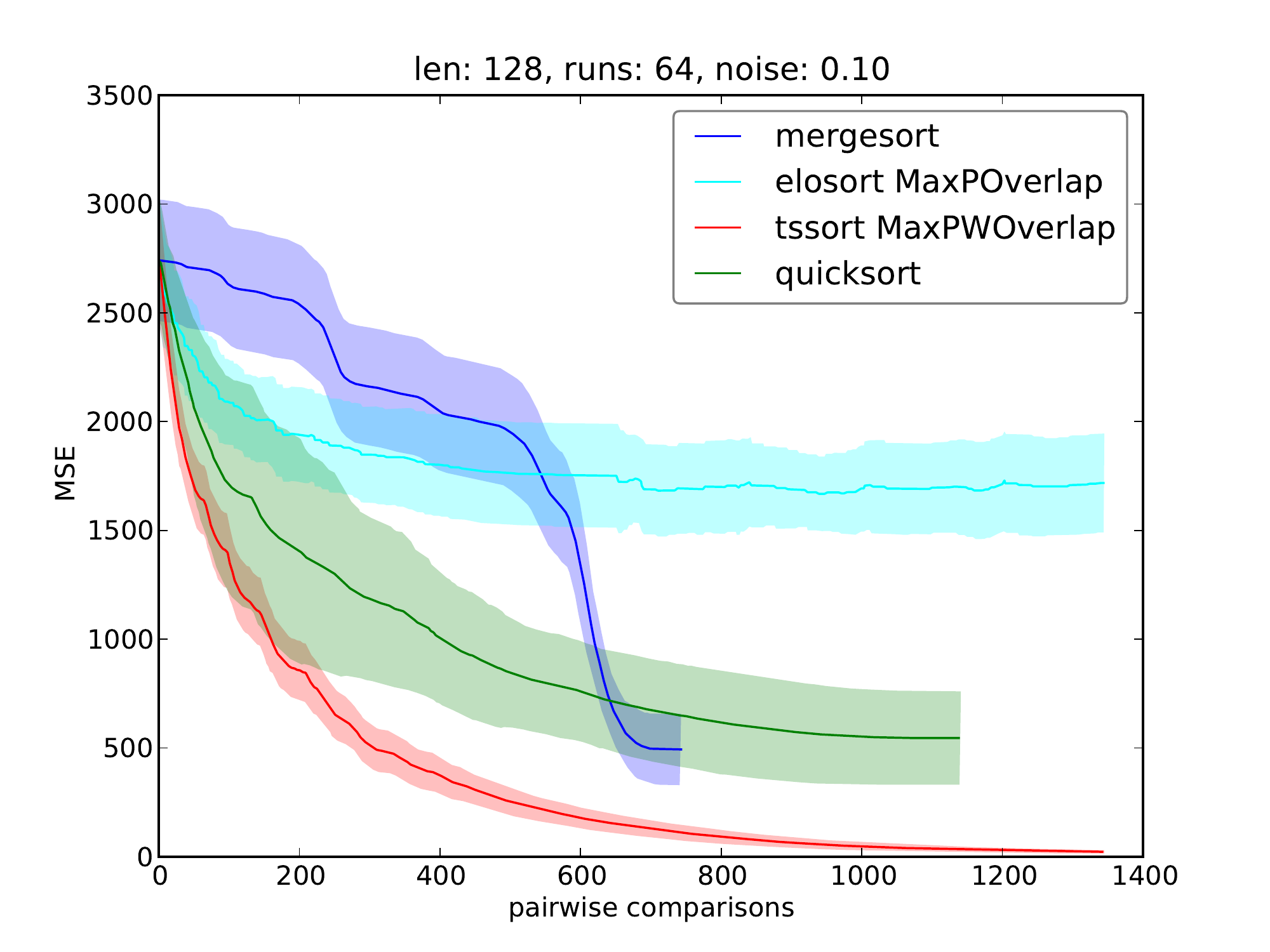} \hfill
		\includegraphics[width=.47\textwidth]{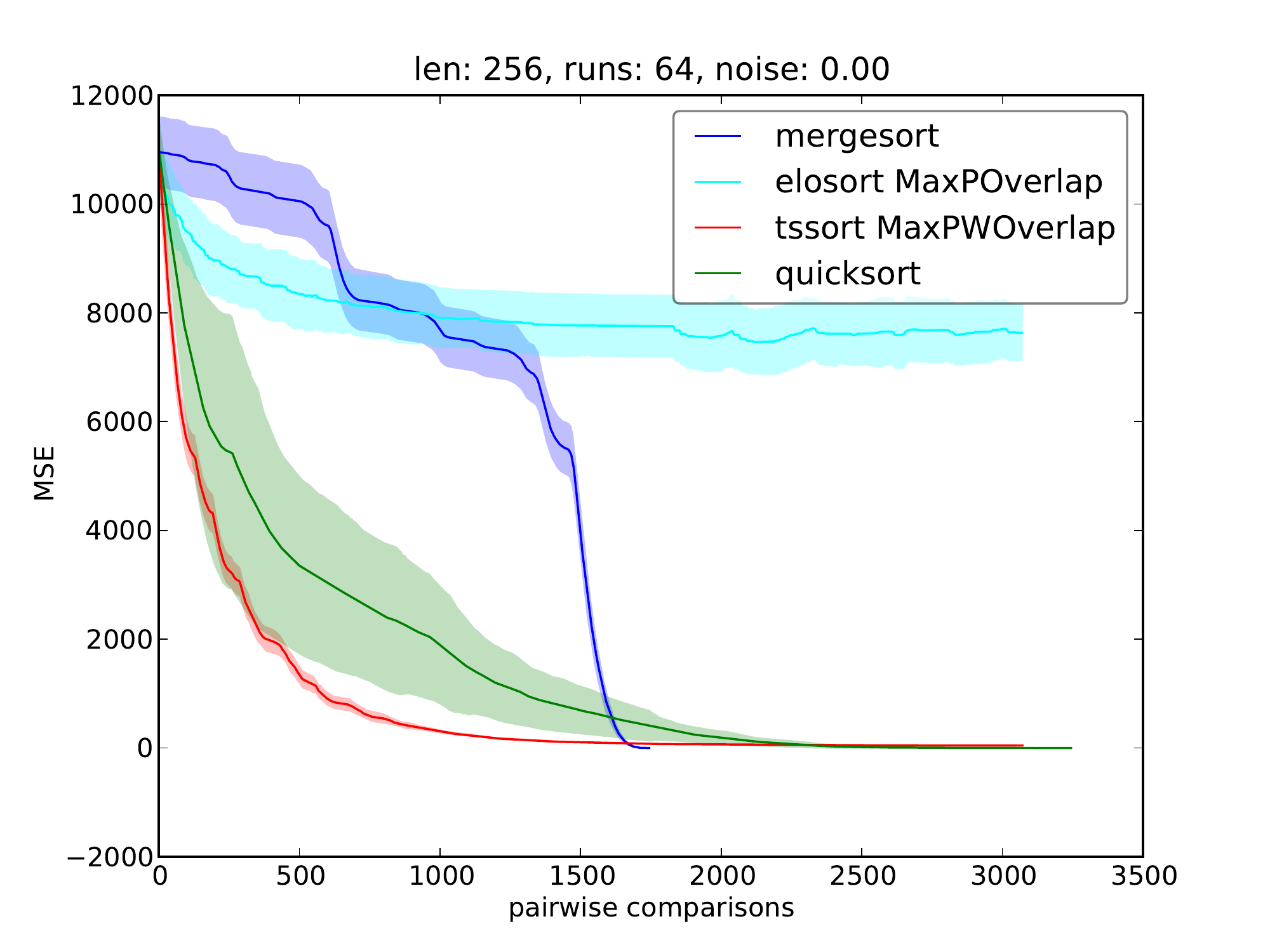} \hfill
		\includegraphics[width=.47\textwidth]{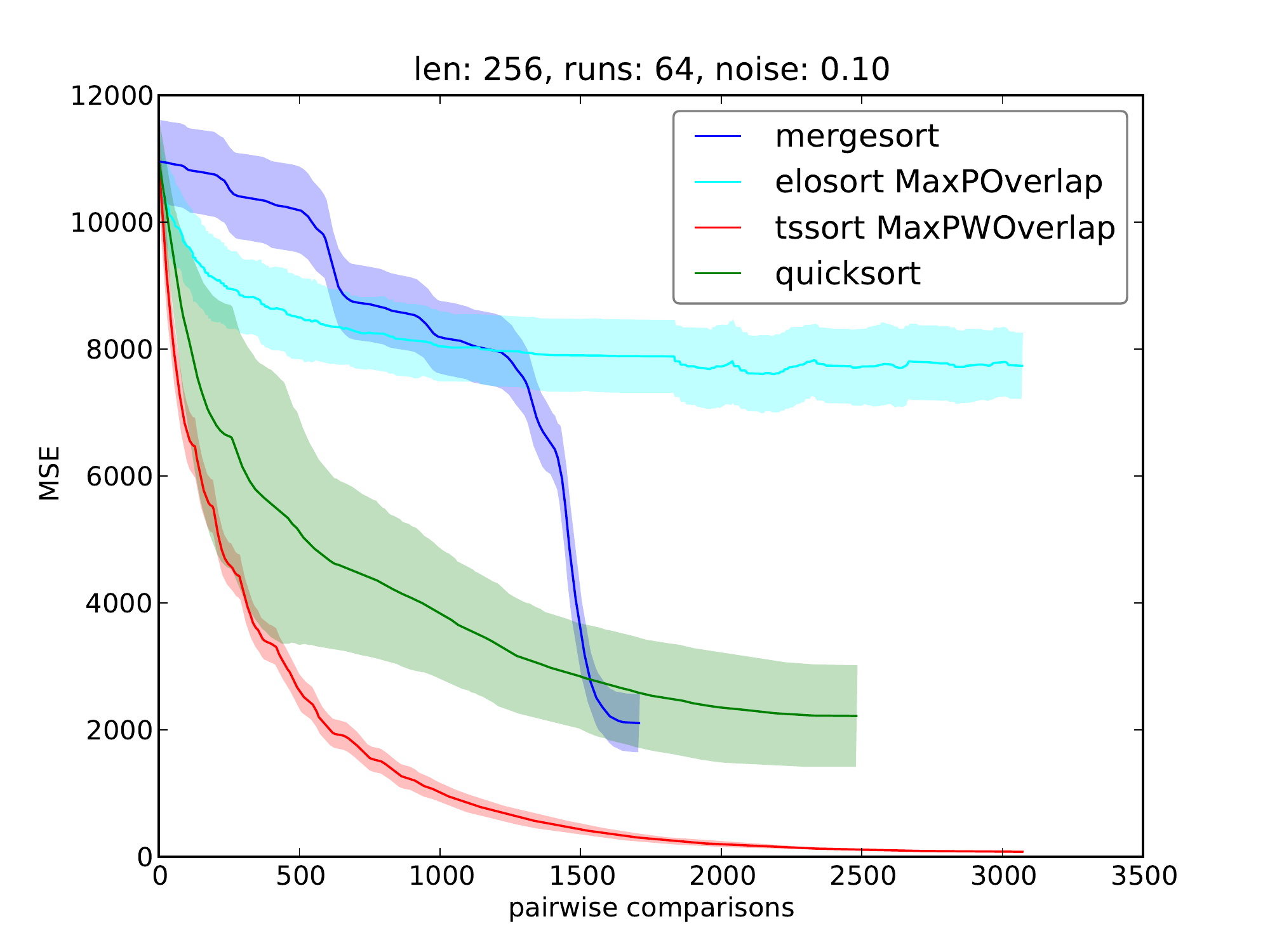} \hfill
		\includegraphics[width=.47\textwidth]{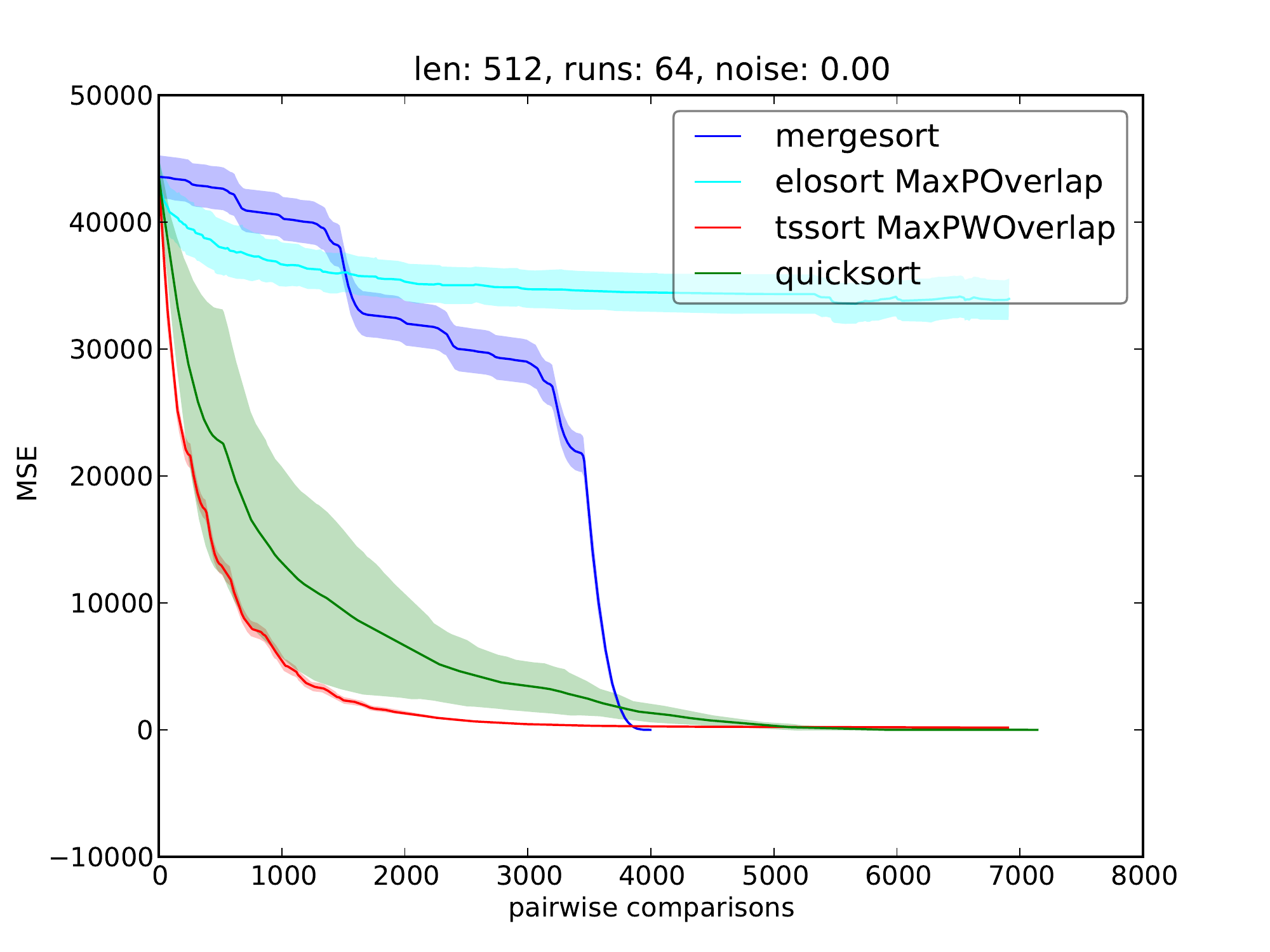} \hfill
		\includegraphics[width=.47\textwidth]{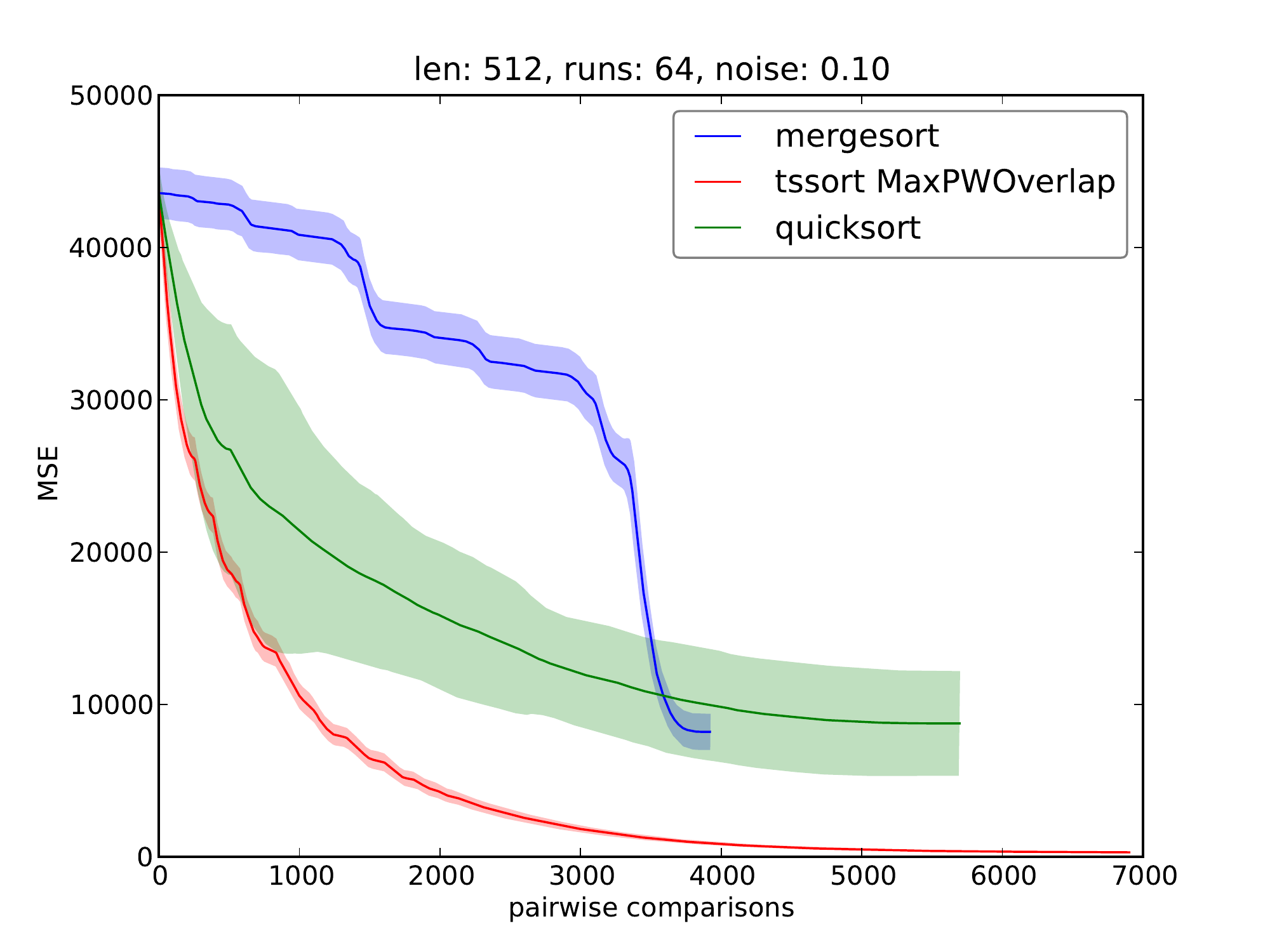} \hfill
		\caption{Comparison of sorting algorithms (Lists of length 128,256,512)}
		\label{fig:compSortAlgos2}
	\end{figure}

	In order to assess the qualities of the investigated sorting algorithms, we ran simulations for lists of several lengths as shown in Figures~\ref{fig:compSortAlgos1} and \ref{fig:compSortAlgos2}.

	Each of the graphics shows the sorting process of a list consisting of numbers from 0 to $n-1$, where $n$ is the length denoted in the title of each graphic (len).
	The left column of graphics shows results for noise level 0 \%.
	In the right columns we simulated a noise level of 10 \% (i.e., with a probability of 10 \% the result of a comparison if $i < j$ was inverted).

	All lists were randomly shuffled then sorted with each of the approaches (each starting from the same randomly shuffled list).
	This process of shuffling and sorting was repeated 128 times for lists with $n\leq64$ and 64 times for longer lists.
	During each sorting run, the intermediate lists were recorded after each comparison of two items.
	For each intermediate step the Mean Square Error (MSE) of the list's items was calculated.
	Over all runs of one sorting algorithm the averages of the MSEs and their respective standard deviations were computed and plotted (y-Axis) after each comparison (x-Axis); the line indicating the average is surrounded by an area showing the standard deviation.

	The compared sorting algorithms include standard comparison based sorting algorithms, such as  bubble-sort (as worst case reference), merge-sort and quick-sort as references.
	Furthermore, the Elo and TrueSkill rating algorithms were converted into one (EloSort) and three (TSSort) different sorting algorithms by combining them with best next item pair selection strategies mentioned in the previous section.

%

	In the noiseless graphics we can find merge-sort to always be the fastest algorithm when it comes to the final sorting time.
	While, as mentioned before, merge-sort shows a rather slow convergence to a sorted list in the beginning, it is unbeatable in the end.
	Quick-sort on the other hand shows faster convergence towards a sorted list in the beginning.
	As expected, all of the traditional sorting algorithms run into problems as soon as the comparison operator gets noisy.

	As mentioned in Section~\ref{sec:relatedWork}, we can see that the Elo rating system with the suggested maximum partner overlap selection strategy does not perform well.
	Even in noiseless cases it is not able to sort the generated lists by comparing selected item pairs.
	The main reason for this is its simplification that all items have the same variance $\beta^2$ of individual performances.
	Also the Elo's updating rule does not seem to be designed for a cold starting phase where all items are set to a rating of 1000.
	After the start, the item ratings subdivide into two concentration points of rating scores ($1000 \pm \Delta$).
	If two items with $s_i,s_j = 1000 + \Delta$ are selected in the next step, the loser's $s$ will afterwards be $1000$ again, which causes many unnecessary repetitions.
	Additionally, the Elo rating system does not include a per item damping term, but instead all items depend on the item independent $K$-factor.

	It is interesting to see that all of the tested TrueSkill based sorting approaches work very well.
	All of them outperform the above mentioned standard approaches when it comes to quick convergence of the intermediate lists.
	Also the TSSort approaches seem to be able to handle noise quite well in contrast to all other approaches.
	A closer look to the approach using the original maximum draw probability selection strategy shows that while being a lot worse than the other proposed selection strategies in the beginning, it catches up in the end in most cases.
	The suggested maximum weighted overlap and the maximum partner weighted overlap strategies show little if any differences allowing us to save a lot computation time by just calculating the maximum partner weighted overlaps for successive pairs.

	Excluding computationally expensive algorithms, we can see that the TrueSkill based probabilistic sorting algorithm with the maximum partner weighted overlap selection strategy for the best next game item pair also performs nicely when it comes to lists of larger size (256 or 512) elements.

	Sadly, aside from its good performance and ability to cope with noise, it was not possible to find a TrueSkill inherent stopping condition, which would allow to dynamically stop sorting the an item list as soon as the algorithm notices that the ordering is sufficient (not necessarily perfect).
	Hence, in the current implementation we propose a limit of $n\cdot \log_2(n)$ updates based on the simulation results.
	After this amount of updates, an item list with $n$ items is assumed to be sufficiently sorted.
	Nevertheless, it remains as future work to investigate if we can find such a dynamic stopping condition. 


\section{Conclusion}
%
In this paper we presented our approaches to create a probabilistic, noise resistant sorting algorithm which converges very quickly towards a well ordered list.
We defined desired properties for such an algorithm:
Minimized waste of decisions, noise resistance and quick convergence.

In our simulations we compared EloSort and TSSort with standard comparison sorts and found TSSort combined with our newly developed Maximum Partner Weighted Overlap selection strategy to be superior to the other algorithms w.r.t. our defined properties.

Even though TSSort shows very good performances, a lot of open questions remain.
Besides evaluating other rating algorithms and tuning TSSort's parameters, it would be interesting to investigate if there is a TSSort inherent stopping condition which allows the algorithm to stop sorting as soon as it considers the list sufficiently ordered.
Apart from this there are many application scenarios which have slightly different requirements than our desired properties.
For example, one might be interested in selection strategies which lead to very precise sortings at the top or bottom end of a list, while being imprecise elsewhere.
Last but not least, in this paper we assumed noise to occur uniformly over all comparisons, but it would also be interesting to model noise to occur more often, the closer the items we compare are together.

\bibliography{library}
\bibliographystyle{unsrt}

\end{document}